# A Clustering-based Location Privacy Protection Scheme for Pervasive Computing

Lin Yao[1, 2], Chi Lin[1], Xiangwei Kong[2], Feng Xia[1], Guowei Wu[1]
[1]School of Software, Dalian University of Technology, Dalian 116620, China
e-mail: f.xia@ieee.org; yaolin_yl@hotmail.com
[2]School of Electronics & Information Engineering, Dalian University of Technology, Dalian 116024, China

*Abstract*—**In pervasive computing environments, Location-Based Services (LBSs) are becoming increasingly important due to continuous advances in mobile networks and positioning technologies. Nevertheless, the wide deployment of LBSs can jeopardize the location privacy of mobile users. Consequently, providing safeguards for location privacy of mobile users against being attacked is an important research issue. In this paper a new scheme for safeguarding location privacy is proposed. Our approach supports location K-anonymity for a wide range of mobile users with their own desired anonymity levels by clustering. The whole area of all users is divided into clusters recursively in order to get the Minimum Bounding Rectangle (MBR). The exact location information of a user is replaced by his MBR. Privacy analysis shows that our approach can achieve high resilience to location privacy threats and provide more privacy than users expect. Complexity analysis shows clusters can be adjusted in real time as mobile users join or leave. Moreover, the clustering algorithms possess strong robustness.**

## I. INTRODUCTION

The proliferation of smart gadgets, applications, mobile devices, PDA and sensors has enabled the construction of pervasive computing environments, transforming regular physical spaces into "Active Information Spaces" augmented with intelligence and enhanced with services [1-3]. Location-Based Services (LBSs) are one of the most desirable classes of services to be offered in pervasive computing environments. Service providers envision offering many new services based on a user's location as well as augmenting many existing services with location information [4]. Considering this scenario, someone wants to have dinner and is searching for a restaurant using the Internet. In order to get more accurate and useful research results, more terms such as the mobile user's location, the type of food, etc. should be included in his search criteria. Unfortunately, if the queries are not securely managed, it could be possible for a third party to retrieve the mobile user's personal sensitive information such as his location information, his habit, etc. In this case, even if an individual does not directly release personal information to the service provider, this provider may become aware of the sensitive information if it has to provide a service to such an individual.

Privacy in pervasive computing environment includes anonymity, context, confidentiality and integrity [5]. Except users who want to disclose their context information (e.g., location, duration, name of service, etc.), no one including outsiders and service providers should know about such information. Location Privacy is a particular type of context privacy. It is defined as the ability to prevent other unauthorized parties from learning one's current or past location. In LBSs, there are two types of location privacy [6]: personal subscriber level privacy and corporate enterprise-level privacy. Personal subscriber-level privacy must supply rights and options to individuals to control when, why, and how their location is used by an application, and to prevent other parties from learning one's past or current location.

Location privacy threats refer to the risks that an adversary can obtain the mobile user's location data. Furthermore, if the LBS provider is unreliable, the location information may be abused and the users may face undesired advertisements, e-coupons, etc. Motivated by this fact, a new method of protecting location privacy based on clustering is developed in this paper. Specially, we prevent an attacker from inferring the real location information of the mobile user by adapting the K-anonymity technique to the spatial domain.

The concept of K-anonymity was introduced as characterizing the degree of data protection with respect to inference by linking [7]. K-anonymity can be ensured in information release by generalizing and/or suppressing part of the data to be disclosed. A data release is said to meet K-anonymity if every topple released cannot be related to fewer than K respondents, where K is a positive integer set by the data holder. In order to protect the location information of mobile users in the context of LBSs, Gruteser and Grunwald [8] firstly employed K-anonymity. A subject is considered as K-anonymity with respect to location information, if and only if the location information sent from one mobile user is indistinguishable from the location information of at least K-1 other mobile users. The spatio-temporal cloaking [8] assumes that all users have the same K-anonymity requirements and K cannot vary with the different privacy requirements of different users. In order to increase the scalability, a customizable K-anonymity model instead of a uniform K was proposed in [9]. Every user can specify a different K-anonymity value based on his minimum anonymity level and his preferred spatial and temporal tolerance level in order to maintain the personalized variable privacy requirements. This model can avoid the drawback of a large K-anonymity spatial region, which is an area that encloses the mobile user querying to a LBS server. However, due to the computation overhead of the clique graph, this approach is only able to meet the small K-anonymity requirements of mobile users.

In order to solve the above drawbacks [9], Casper [10] is proposed. Casper includes a location anonymizer and a privacy-aware query processor. The location anonymizer implements the location K-anonymity based on the specified privacy requirements. The privacy-aware query processor deals with the cloaked spatial areas rather than the exact location information. Though Casper can achieve high quality LBSs, it cannot meet the QoS requirements of mobile users. Therefore an efficient message perturbation engine [11] is developed. On one hand, the message perturbation engine can effectively implement location K-anonymity. On the other hand, the QoS requirements can be met. But only a part of users can get the ideal levels of privacy as they require. Some of the users' requests cannot be delivered to the LBS providers permanently because they may be missed in the spatial cloaking algorithm. Neighbor-k and local-k methods are used to establish the cloaking region, which can only achieve the local optimal resolution without getting the global resolution. The spatial and temporal resolution is realized at the expense of denying several users' requests.

In this paper, clustering algorithms are used to tackle such drawbacks of the cloaking algorithm mentioned in [11]. Our proposed system consists of a trusted third party (TTP) acting as a middle layer between mobile users and LBS providers. First, TTP receives the exact location information from a mobile user. Then, TTP will blur the exact information into a cloaked spatial area using clustering algorithms. Next, a list of results will be sent to TTP because the LBS provider cannot receive the exact location information but the cloaked area. Finally, TTP will select the most optimal result to the mobile user from the list. Therefore, a mobile user can enjoy LBSs and get more privacy without revealing his private location information.

The rest of this paper is organized as follows. In section 2, we describe the system architecture. Section 3 presents the clustering algorithms in detail. The system theoretical analysis is given in section 4. Section 5 shows experiment results. Finally, we conclude the paper in section 6.

## II. Problem Formulation

In this section, we describe the architecture of our location privacy protection system in Fig.1. Mobile users communicate with LBS providers through TTP. TTP acts as an anonymity server between mobile users and LBS providers. Cluster algorithms are running in TTP, and TTP helps to achieve location K-anonymity according to the required anonymity level of each mobile user.

Supposing that the channel between every mobile user and TTP is secure, the exact location information from a mobile user to TTP can be protected from being obtained by attackers. TTP blurs the decrypted exact location information into a cloaked spatial area with clustering algorithms. The cloaked area composed of K users is sent to the LBS provider. Due to lack of the mobile user's exact location information, the LBS provider may send back a list of results to TTP. Lastly, TTP will select the most optimal result to the mobile user based on his exact location information. The value of K can vary with the anonymity level of each mobile user. The following six steps in Fig. 1 describe the whole process.

(1) Every mobile user sends a message consisting of his exact location, K, and a LBS request.

(2) All users are clustered as soon as TTP receives the message.

(3) The exact location information is replaced by MBR of the cluster where the mobile user locates.

(4) LBS returns a list of results to TTP in light of the received MBR from step 3.

(5) TTP sends the optimal result to the mobile user based on the exact location information in step 1.

(6) The mobile user receives the result from step 5.

Since a mobile device possesses limited memory and limited computing capabilities, it cannot act as an anonymity server instead of TTP.

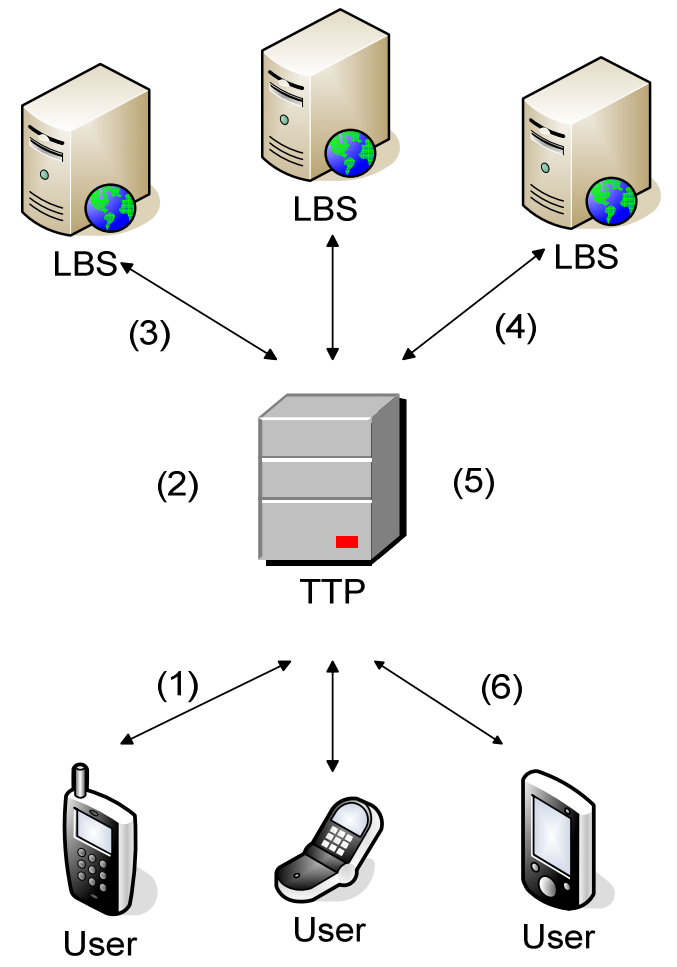

Figure 1. System Architecture

## III. Clustering Algorithms

In this work we propose a location privacy protection scheme called ClusterCloak for the above described pervasive computing environment. ClusterCloak is run by TTP to blur a user's exact information. When a mobile user requests LBS, he will send a user profile to TTP. A user profile is a message defined as follows: $m_s \in S:\{u_{id}, n_{id}, (x,y), K, Ct\}$. The payload content in ms is omitted. The notions are listed in Table I.

As soon as receiving $m_s$, TTP divides the whole area into several clusters. The exact location information in $m_s$ is replaced by MBR of the user's cluster so as to achieve K-anonymity. Consequently TTP sends a message $m_t$ to LBS. Let $\phi(t,s)=[t-s,t+s]$, which extends a numerical value $t$ to a range by amount $s$. $m_t$ is defined as follows.

$$m_t \in T:\{u_{id}, n_{id}, X:\phi(cx,\frac{1}{2}W_{MBR}),\ Y:\phi(cy,\frac{1}{2}H_{MBR}), C\,t\}$$

The ClusterCloak has the following features:

1) *K-anonymity*: K in $m_s$ means the anonymity level the user desires, which must be met by TTP.

2) *High quality QoS*: There is a balance point between QoS and K in the past approach [11]. High quality QoS requires the diminution of the cloaked area, which may increase the chance of being found by attackers. Conversely, the extension of the cloaked area may influence the accuracy of the query results and make the user not attain the optimal result. In our paper, ClusterCloak can solve the above contradictoriness. Adjacent users can be divided into the same cluster; hence the cloaked area may dwindle as far as possible on the premise of meeting the required K-anonymity level.

3) *High efficiency*: The mobile user's query must be responded by the LBS provider in real time. Even if users move, TTP must finish clusters adjustment quickly.

4) *Robustness*: In some cases, though users move, clusters do not need adjusting, which reduces the workload of TTP.

The recursive cluster algorithms will be introduced in detail in the following.

TABLE I. LIST OF NOTATIONS

| Notation | Description |
|---|---|
| $S$ | A message set the source sends |
| $T$ | A message set TTP sends |
| $m_s$ | A message in set $S$ |
| $m_t$ | A message in set $T$ |
| $u_{id}$ | User ID |
| $n_{id}$ | Message ID |
| $K$ | Anonymity level |
| $cx$, $cy$ | Coordinate of center of every cluster |
| $x$,$y$ | Coordinate of a user |
| $X$,$Y$ | Coordinate range of MBR |
| $H_{MBR}$ | Height of MBR |
| $W_{MBR}$ | Width of MBR |
| $Ct$ | Content of message |
| $c_i$ | The *i*-th cluster |

### *3.1 Building Clusters*

ClusterCloak is composed of six algorithms. Next we will introduce the algorithms in detail.

#### *3.1.1 Related Definitions*

*Definition 1: Cluster Area* is a circle whose radius is the distance from the center to the remotest point in the cluster.

*Definition 2: Neighbor Clusters* are two tangent clusters or two intersecting clusters.

*Definition 3:* $P_{need}$ is the probability of rebuilding a cluster when a mobile user moves.

*Definition 4:* $N_{ex}$ is the number of extra nodes, without which the cluster can still keep robust.

*Cluster Area* and *Neighbor Clusters* are used in the process of cluster merging. $P_{need}$ and $N_{ex}$ are utilized to judge if a cluster needs dividing.

#### *3.1.2 The Choice of Initial Center*

The choice of initial center has strong relations with the complexity of building clusters. In this paper, the following four methods are adopted.

*MN*: Two nearest points along the vertical direction or the horizontal direction in the MBR are selected.

*NR*: One point is selected randomly, and the other is the nearest one to this point.

*RP*: Two points are selected randomly.

*RS*: All points are divided into two sets in the horizontal way, and a random point is selected as the center of every set.

#### *3.1.3 Building Clusters*

After selecting the cluster center, each point is assigned to the nearest cluster according to the distance from it to the center. Then new center will be calculated and each point is assigned to the nearest cluster again. The above process will repeat until the sum distance between every point and cluster center (CDS) converges to a certain range. The new center and CDS are defined as follows:

$$cx = \frac{1}{\|C_i\|}\sum_{j\in C_i} x_j$$

$$cy = \frac{1}{\|C_i\|}\sum_{j\in C_i} y_j$$

($x_j$, $y_j$) is the coordinate of a point $j$ in the cluster $C_i$.

$$CDS = \sum_{j\in C_i}\sqrt{(x_j - cx)^2 + (y_j - cy)^2}$$

We use $c_i.CDS$ to stand for the CDS of *i*-th cluster.

The process of building clusters is illustrated in Algorithm 1 which applies a bipartite cluster method. The initial phase is in lines 2-4. *Cm* is defined as a structure which records the cluster identifier (ID), the nodes ID, the cluster center, the cluster size, CDS, MBR, $P_{need}$, $N_{ex}$, and a variable *divided*. The local variable *divided* represents if the cluster needs dividing. Its value is true or false, which depends on the value of $P_{need}$ and $N_{ex}$. When $P_{need}$ is equal to zero and $N_{ex}$ is more than one, *divided* is true. Otherwise, *divided* is false meaning the cluster needs converting.

*Cm* changes as a new cluster is created or an old cluster is merged. Initially, *Cm* only contains the initial cluster $c_0$, so Algorithm 2 is called recursively in lines 5-11.

In Algorithm 2, if an old cluster $c_j$ can be divided into two new clusters *ca* and *cb*, then $c_j$ is deleted and *ca* and *cb* are inserted into *Cm*. The phase of partitioning a cluster is in lines 3-15. When a user leaves the old cluster and joins another one, both centers will be adjusted.

**Algorithm 1** Building clusters

1 **Function** BuildingCluster
2 // Initially, the whole area is seen as a big cluster $c_0$.
3 $c_0.CDS = \infty$ .//Initialize the CDS of $c_0$
4 Insert $c_0$ into *Cm*.
5 **While** (true)
6     **If** $\forall c_i \in Cm, \exists c_j.divided \neq false$ **Then**
7       **If** $c_j.divided = true$ **Then**
8         Generate two initial center points: *va*, *vb*.
9         *BinaryCluster($c_j$,va,vb)*
10     **Else** break
11 **End**

**Algorithm 2** Cluster division

**Function** BinaryCluster($c_j$, $va$, $vb$)
//Two new clusters $ca$ and $cb$ are initialized
**Repeat**
**For each** $p_i \in c_j$
  **If** *distance*($p_i$,*va*)>*distance*($p_i$,*vb*) **Then**
    $p_i$ .*clusterID* = *va* .*clusterID*
    *ca.add*( $p_i$ )
  **Else** $p_i$ .*clusterID* = *vb.dlusterID*
    *cb.add*( $p_i$ )
**End**
Re-calculate center of *ca, cb*
**If** $ca.P_{need} = -1 \vee cb.P_{need} = -1$ **Then**
  $c_j$.*divided* = -1
  **Return** FALSE // $c_j$ cannot be divided.
**Until** the center of *ca* and *cb* do not change.
Insert *ca*, *cb* into *Cm*.
Delete $c_j$.
**Return** TRUE

*3.2 Adjusting Clusters*

In pervasive computing environments, a mobile user is roaming from one domain to another domain, so clusters may be adjusted. Firstly, a cluster does not need adjusting if a user roams in his original cluster. Secondly, a user will be assigned to the nearest cluster if he leaves his home cluster. If his home cluster cannot meet the K-anonymity level, it should be merged with its nearest cluster.

*3.2.1 A User's Joining*

We denote $k_1$, $k_2$, …, $k_m$ as anonymity levels of $m$ users, where $k_1$, $k_2$, …, $k_m$ are arranged in the ascending order. When one or multi-users join a new cluster, Algorithm 3 will be implemented to divide the cluster into two clusters. But if either cluster cannot meet the requirement of K-anonymity, cluster adjustment is not successful. Only $P_{need}$ and $N_{ex}$ are re-calculated and users can obtain higher privacy levels because the cluster size is larger than $k_m$. Algorithm 4 is proposed to adjust the cluster whenever the size of the cluster has changed.

**Algorithm 3** A user's partition

**Function** PointInsertion($p$)
Find the nearest cluster $c_i$ .
$c_i$ *.add(p).*
Update $c_i$ in *Cm.*
**If** $0 < c_i.P_{need} \le 1$ **Then**
  $c_i.P_{need} = 1, c_i.N_{ex} = 1$ .
**Else** $c_i.N_{ex} = c_i.N_{ex} + 1$ .
*ClusterAdjustment(* $c_i$ *).*

**Algorithm 4** A cluster's adjusting

**Function** ClusterAdjustment($c_i$)
$c_i$ *.divided = true.*
Adjust the *CDS* of $c_i$.
Same as lines 4-10 in Algorithm 1 to iteratively divide $c_i$.

*3.2.2 A User's Leaving*

When a user leaves the home cluster, four scenarios may occur. Algorithm 5 illustrates the adjustment of cluster when a user leaves.

Firstly, if $m$ is bigger than $k_m$, the cluster can still keep robust. When one user leaves, it holds that $m - 1 \ge k_m$, which means the anonymity levels of the rest users can still be met. Therefore, $P_{need}$ and $N_{ex}$ are re-calculated and the cluster does not need rebuilding.

Secondly, if the equations of $m = k_m$ and $k_m > k_{m-1}$ are true, a user whose anonymity level is $k_m$ leaves the cluster. The cluster size will become $m$-1, and $k_1 \le k_2 \le \ldots \le k_{m-1} \le m$-1 is true. The anonymity levels of the rest users can still be met. Therefore the cluster does not need rebuilding.

Thirdly, if the equation of $m = k_m$ is true, a user whose anonymity level is $k_i$ leaves ($k_i \neq k_m$). Therefore, the cluster size will become $m$-1. $k_m$ is bigger than $m$-1. Therefore the cluster needs merging and Algorithm 6 is called.

It can be drawn from the second and third scenarios that

$$P_{need} = \frac{m-1}{m}.$$

Last, if the equations of $m = k_m$ and $k_m = k_{m-1}$ are true, any user in the cluster leaves. The cluster size becomes $m$-1. The anonymity level of $k_m$ or $k_{m-1}$ cannot be met. Algorithm 6 will be implemented to rebuild clusters and hence $P_{need}$ = 1.

**Algorithm 5** A user's leaving

**Function** PointQuit($p$)
Find the cluster $c_i$ in which $p$ resides.
**If** $c_i$ . $N_{ex}$ > 1 **Then**
  $c_i$ *.del(p).*
  *ClusterAdjustment(* $c_i$ *).*
**Else If** $c_i$ . $N_{ex}$ = 1 **Then**
  $c_i$ . $N_{ex}$ = *0.*
  Adjust $c_i$ . $P_{need}$ = 1.
**Else** *ClusterMerge(ci).*
Update *Cm.*

*3.2.3 Clusters Merging*

When K-anonymity cannot be met because of a user's leaving, the cluster should be merged with a neighbor that owns the minimum MBR. Algorithm 6 will be implemented.

**Algorithm 6** Clusters mergence and division

1 **Function** ClusterMerge($c_i$)
2 Record the *Neighbor Clusters* of $c_i$ with the largest $N_{ex}$ in *MCm* .
3 **If** $\|MCm\| \ge 1$ **Then** // Size of *MCm* is bigger than 1
4   Select the cluster $c_j$ with minimum *MBR.*
5 **Foreach** $p_s \in c_i$
6   $c_j$.*add*($p_s$).
7 **End**
8 **Delete** $c_i$
9 *ClusterAdjustment(* $c_j$ *).*

In line 2 of Algorithm 6, TTP searches the neighbor cluster of $c_i$ with the minimum MBR. Users in $c_i$ will be added into the neighbor cluster $c_j$, and then $c_i$ is deleted from $Cm$. At last, Algorithm 4 is called to divide $c_j$ into smaller ones.

## IV. SYSTEM ANALYSIS

In this section, ClusterCloak will be analyzed theoretically in terms of privacy and performance.

### *4.1 Privacy Analysis*

ClusterCloak aims to protect location privacy with personalized K-anonymity. K-anonymity represents that the attacked probability of each user is $1/K$ in a region of $K$ users. For any cluster, $\|C\|$ is defined as the number of users in the cluster and $k_m$ is defined as the maximum K-anonymity level. In our ClusterCloak, each cluster is built based on $\|C\| \geq k_m$ which indicates that any person can get more privacy than he expects. $R_k = \|C\| / k$ is defined as the relative K-anonymity. Therefore, the bigger $R_k$ is, the more privacy TTP can provide. In Fig. 2, $R_k$ is illustrated. MN can provide the highest $R_k$, but it cannot provide a constant $R_k$. MN shows an ascending trend during the interval [100-400] and tends to balance at last. However, NR, RP and RS remain a constant $R_k$ during the whole process. In summary, all methods can guarantee $R_k > 1$ so as to provide higher anonymity levels than users expect.

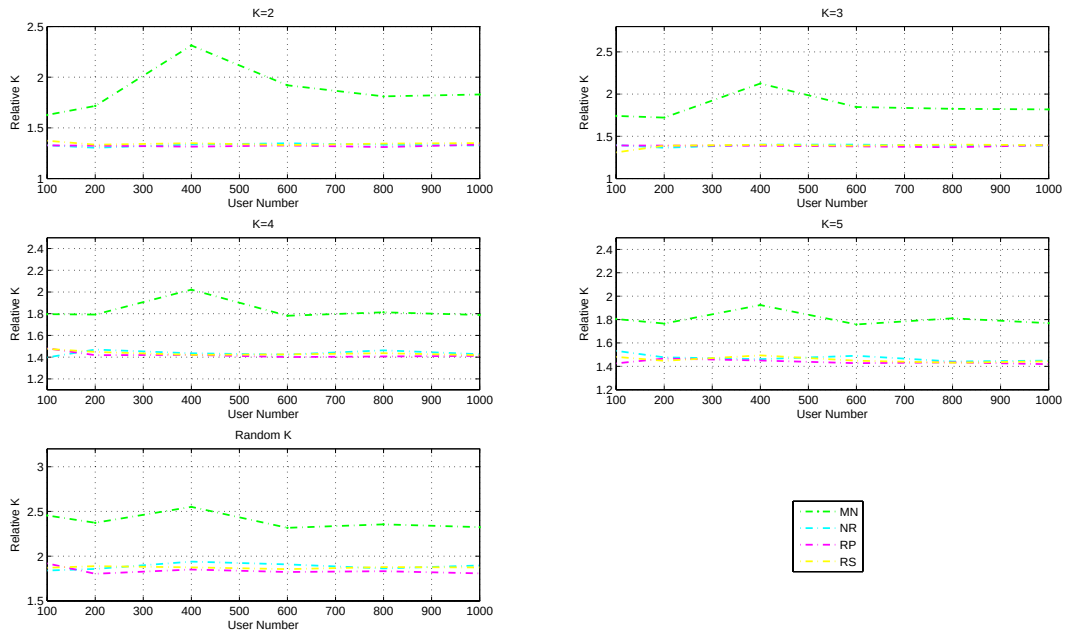

Figure 2. Relative K-anonymity

To further prove that ClusterCloak can provide higher privacy level, entropy analysis will be discussed.

Entropy is a concept of information maintaining great importance in physics, chemistry, and information theory. In essence, the most general interpretation of entropy is as a measure of our uncertainty about a system. Greater entropy means more uncertainty.

Let $p_i$ denote the probability that the $i$-th user may be regarded as a target user $T$ by attackers. The entropy of all users is $H(p) = -\sum_{i=1}^{\|C\|} p_i \log_2 p_i$ . Since it can be obtained that $\|C\| \geq k_m$ , $p_1 = p_2 = ...... = p_m = \frac{1}{\|C\|} \leq \frac{1}{k_m}$ and we can hold that $H(p) = \log_2 \|C\| \geq \log_2 k_m$ . In Fig. 3, it also can be seen that the entropy of each method is higher, which indicates that ClusterCloak can provide more uncertainty so as to reduce the chance of being identified by attackers.

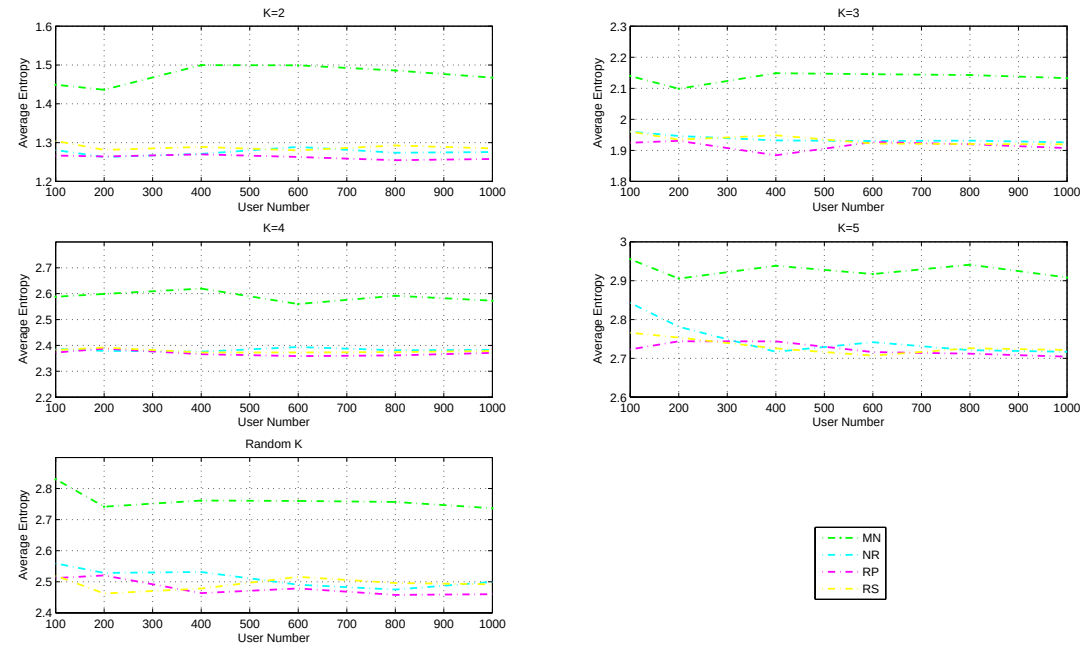

Figure 3. Entropy Analysis

### *4.2 Complexity Analysis of Building Original Clusters*

In order to achieve location K-anonymity, a cloaked area instead of a user's exact location information is sent to LBS providers by TTP. ARNN [13] forms a pyramid structure to search k nearest neighbors to form the cloaked area. Nbr-k and local-k [12] are proposed to make TTP act as a message perturbation engine to form the MBR for the users. HilbertCloak [8] utilizes a Hilbert space filling chain to define a total order among users' locations. By using the curve, HilbertCloak calculates the area to replace the exact location of the users. A Casper [10] proposes a grid structure, and performs a bottom up way to search for the cloaked area. All the above methods can only get local optimal solution, while ClusterCloak can get global optimal solution.

For a cluster containing $n$ users, the complexity of one clustering procedure in ClusterCloak method is $O(nt)$, which can be simplified to $O(n)$, since the number of iterations, i.e. $t$, is constant. In the worst case, the complexity of the recursion process is $T(n) = 2T(n/2) + O(n)$ . Since $O(n)$ is the complexity of each procedure, there must exist a constant $a$ satisfying $T(n) \leq 2T(n/2) + an$ . A full binary tree of complexity can be formed in Fig. 4. The tree height is $lgn$-1. The total complexity is $O(n \times lgn)$.

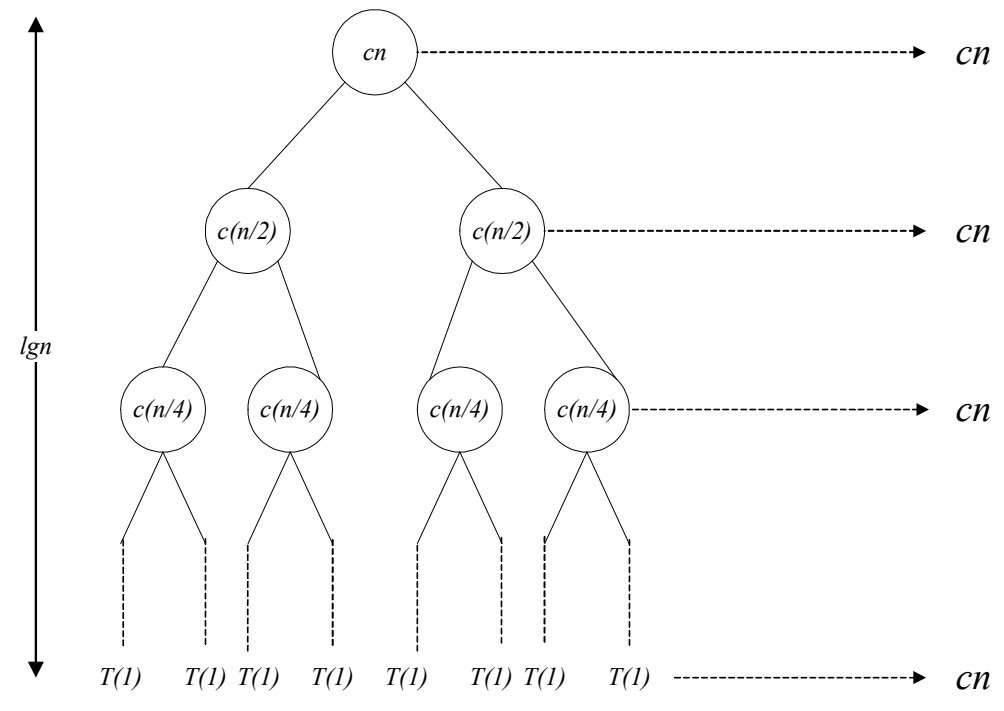

Figure 4. Iteration Tree

As shown in Table II, the complexity of ClusterCloak is $O(n \times lgn)$, which is lower than that of Nbr-k, Local-k and ARNN.

TABLE II. COMPLEXITY OF BUILDING CLUSTERS

| Algorithm | Complexity |
|---|---|
| Nbr-k [12] | $O(n^2)$ |
| Local-k [12] | $O(n^2)$ |
| ARNN [13] | $O(n^2)$ |
| ClusterCloak | $O(n \times lgn)$ |
| Casper [10] | $O(n \times lgn)$ |
| HilbertCloak [8] | $O(n \times lgn)$ |

### *4.3 Complexity Analysis of Adjusting Clusters*

When a user joins or leaves a cluster, the clusters adjustment will be performed. Different from other methods in [8][10][12][13], ClusterCloak can get global optimal solution instead of locally optimal solution. Moreover, the newly cloaked area is adjusted locally when a user moves.

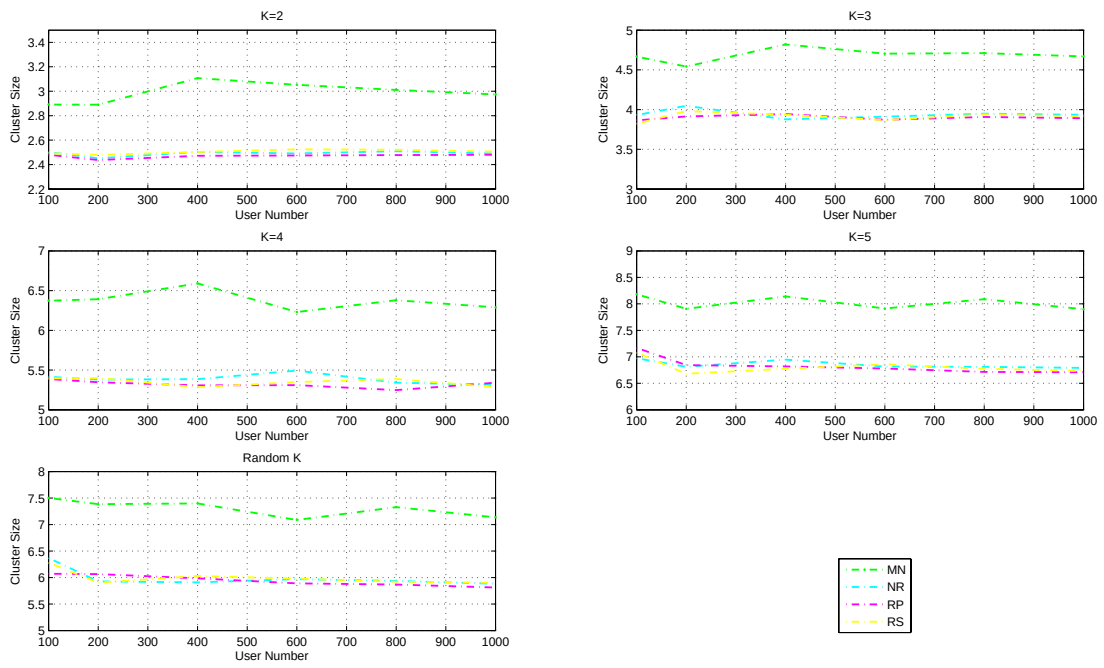

Figure 5. Average Cluster Size

As discussed in Section 4.2, the complexity of one cluster constitution is $O(n)$, which directly depends on the cluster size. Fig.5 illustrates the relationship between the cluster size and the total number of users. It indicates that the cluster size is almost stable. MN is higher than the other methods, but still less than $2k_m$. Therefore, the cluster size can be regarded as a constant.

When a user joins, the cluster size is $\|C\|$ +1. Once a member leaves, the cluster size is $\|C\|$ -1. If two clusters merge, the maximum size is $2\|C\|$ -1 at most. From the above analysis, the complexity of adjusting a cluster is $O(1)$, which reduces the computation cost greatly. The comparison with other algorithms is listed in Table III, which shows that our ClusterCloak has the lowest complexity.

TABLE III. COMPLEXITY OF ADJUSTING CLUSTERS

| Algorithm | Complexity |
|---|---|
| Nbr-k [12] | $O(n^2)$ |
| Local-k [12] | $O(n^2)$ |
| ARNN [13] | $O(n^2)$ |
| ClusterCloak | $O(1)$ |
| Casper [10] | $O(\lg n)$ |
| HilbertCloak [8] | $O(\lg n)$ |

To maintain K-anonymity, Nbr-k, Local-k, and ARNN re-compute a cloaked area when a user leaves or joins a cluster. Casper maintains a quadtree. HilbertCloak maintains a Hilbert chain. Nbr-k, Local-k, and ARNN re-establish the system structure again. In Casper and HilbertCloak, they re-establish the structure, but the cost is up to $O(\lg n)$. Compared with ClusterCloak, the complexity of other algorithms is relatively high.

### *4.4 Complexity of Initial Center*

In this section, we will mainly analyze the complexity of choosing the initial center, which is a critical issue for building clusters. In MN method, at most n points locating in the MBR are traversed in order to generate the two nearest points, so the complexity is $O(n)$. In NR method, a random point is generated at the cost of $O(1)$. The nearest node around the center must be scanned, so n points are compared in the worst case. The complexity is $O(n)$. Two random points are selected in RP method, so its complexity is $O(1)$. In RS method, horizontal dimensions of the users are sorted in an ascending order. Then, the whole region is divided into two equal parts and a random point is selected in every part. If the quick sort is called, the whole complexity is $O(n\lg n)$. The complexity of each method is illustrated in Table IV. It can be seen that none complexity is higher than $O(n\lg n)$.

TABLE IV. COMPLEXITY OF CHOOSING INITIAL CENTER

| Method | Complexity |
|---|---|
| MN | $O(n)$ |
| NR | $O(n)$ |
| RP | $O(1)$ |
| RS | $O(n\lg n)$ |

## V. EXPERIMENTAL ANALYSIS

### *5.1 Experiment Setup*

We use the VANET (Vehicular Ad-Hoc Network) [14] System that simulates movement of cars and generates requests using the position information. Random Map Generator has been performed to create the geographical distribution of the map (Fig.6) and the trace of the vehicles (Fig.7) respectively.

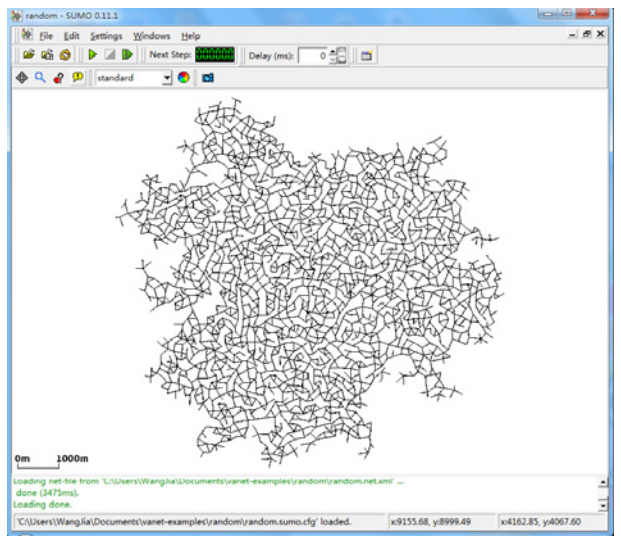

Figure 6. Geographical Distribution

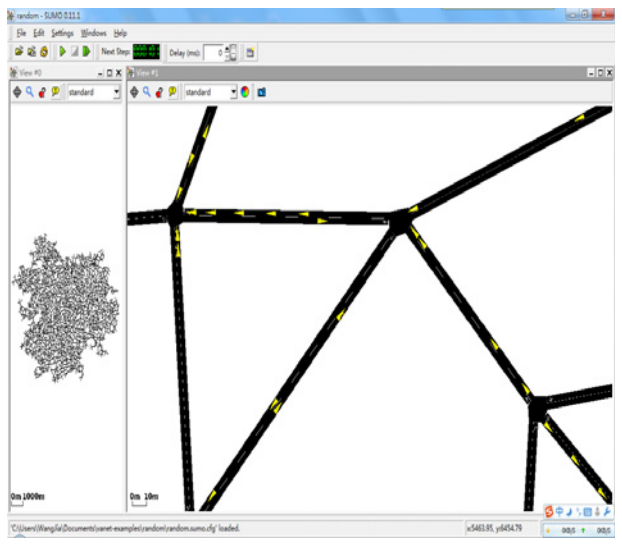

Figure 7. Visualization of Vehicle Trace

The number of mobile users is selected from the list {100, 200, 400, 600, 800, 1000}. The K-anonymity level is from 2 to 5. For every combination of the different number and K, 10 data sets recording users' location information are used. Accordingly the sum of sets is 6×4×10. In order to verify the universality of ClusterCloak, experiments are made when K is randomly selected from 2 to 5. The sum of sets is 6×1×10.

### *5.2 Time Consumption*

In this section, time consumption of building original clusters and adjusting clusters is analyzed.

#### *5.2.1 Building the Original Clusters*

In Fig.8, the relationship between the time consumption of building the initial clusters and the number of random users is depicted. ClusterCloak can finish building clusters in 3s, which is much lower than 5s in [12]. Fig.8 also indicates that different methods such as MN, NR, RP and RS nearly take the same time to establish clusters, which is in accordance with our conclusion in Section 4.4.

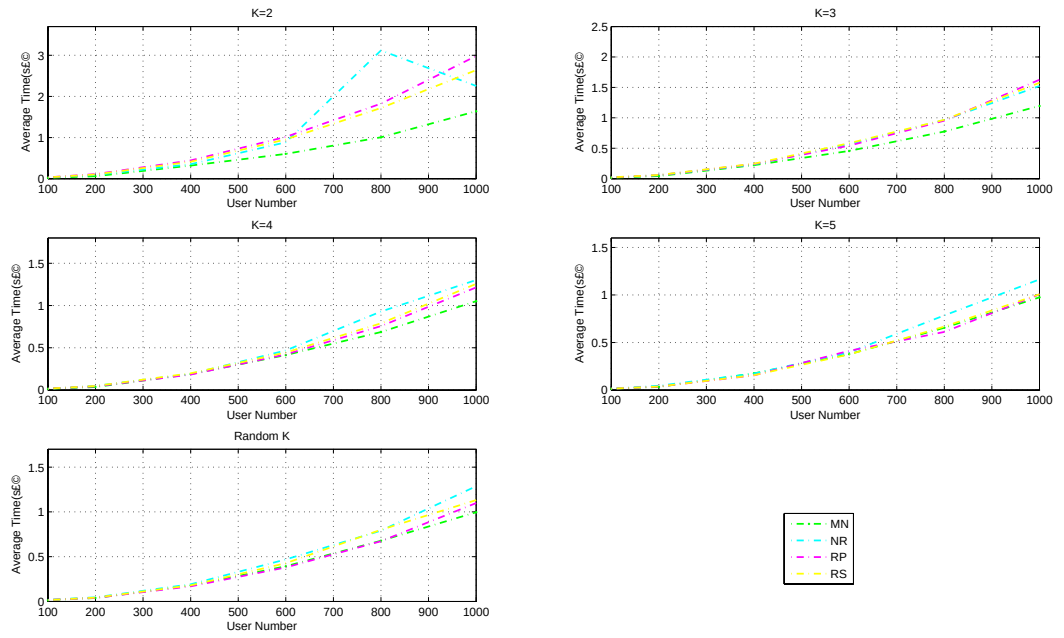

Figure 8. Time Consumption of Building Clusters

#### *5.2.2 Users Joining*

In Fig.9, the relationship between the time consumption of adjusting clusters and the percentage of joining users is depicted. In our experiments, the percentages are 5%, 10%, 15%, and 20% respectively. It can be seen that MN is the fastest and MR is the lowest. The lowest time is 0.15s, which explains ClusterCloak can finish adjusting in no more than 0.15s.

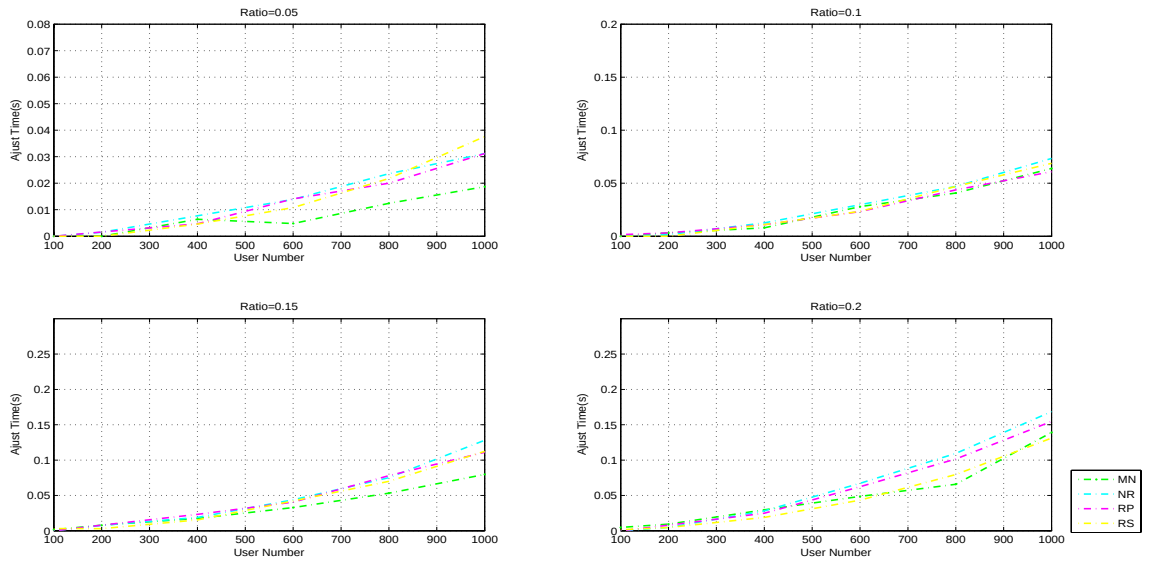

Figure 9. Time Consumption of Users Joining

#### *5.2.3 Users Leaving*

In Fig.10, the relationship between the time consumption of adjusting and the percentage of leaving users is depicted. In our experiments, the percentages are 5%, 10%, 15%, and 20% respectively. It can be seen that the time cost of four methods is almost the same, and the maximum is less than 0.03s. So we can summarize that ClusterCloak can finish adjusting in less than 0.03s.

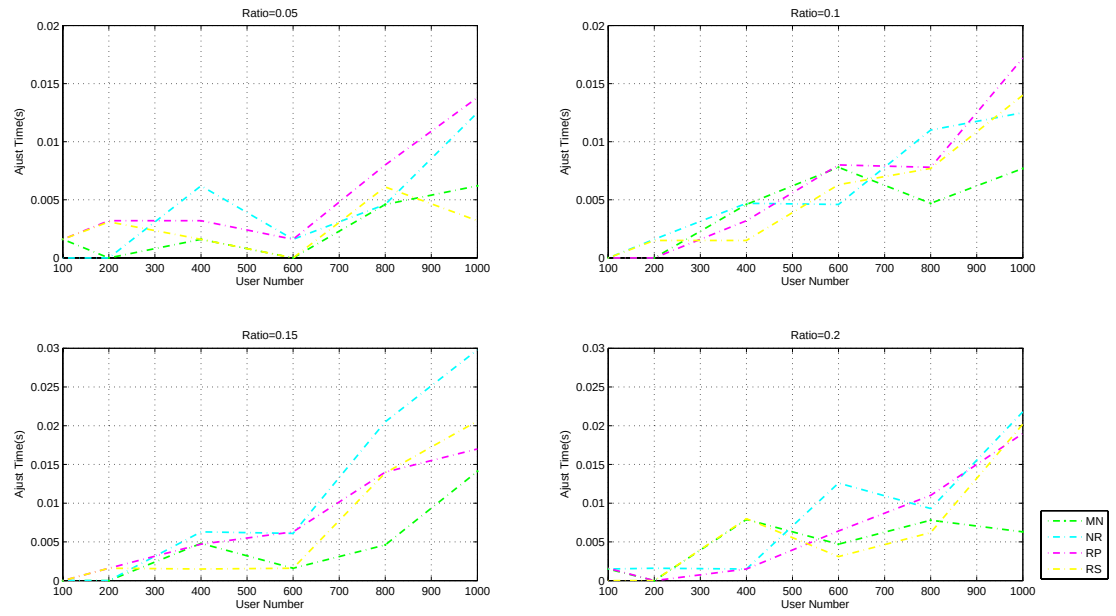

Figure 10. Time Consumption of Users Leaving

### *5.3 Characteristic Analysis of Clusters*

#### *5.3.1 Cluster Stability*

In Fig.11, the relationship between the number of clusters and the number of users is depicted. The number of clusters is linear with the number of users. Though the number of clusters in these methods is different, the slope of each method is constant, which indicates the size of the cluster is nearly a constant. Hence we can draw the conclusion that the size of clusters does not change with the increment of the number of users, which shows that ClusterCloak can build clusters stably.

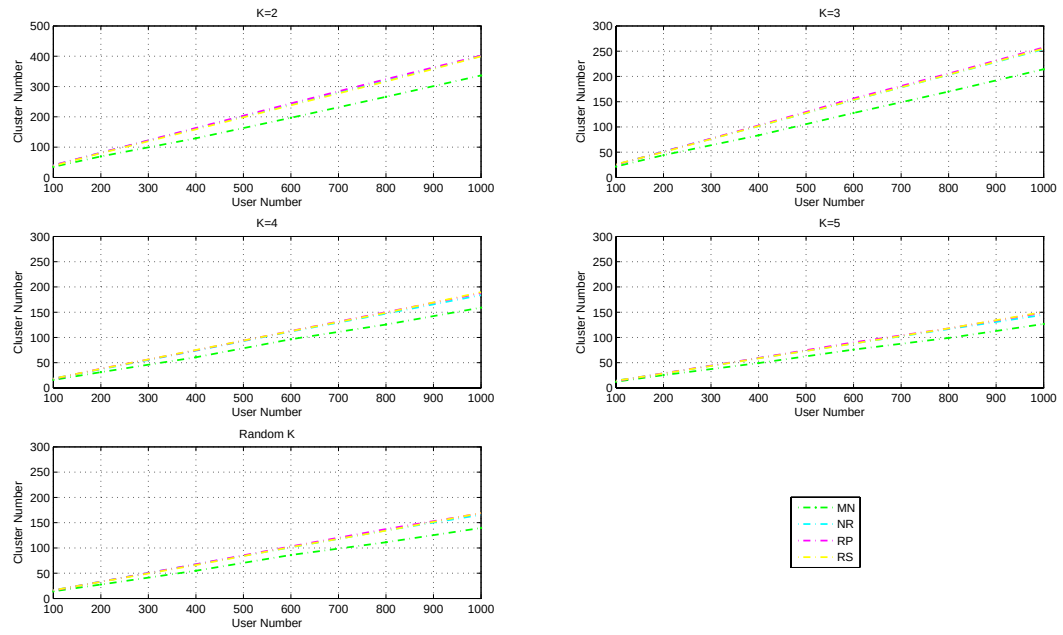

Figure 11. Cluster Stability

#### *5.3.2 Cluster Robustness*

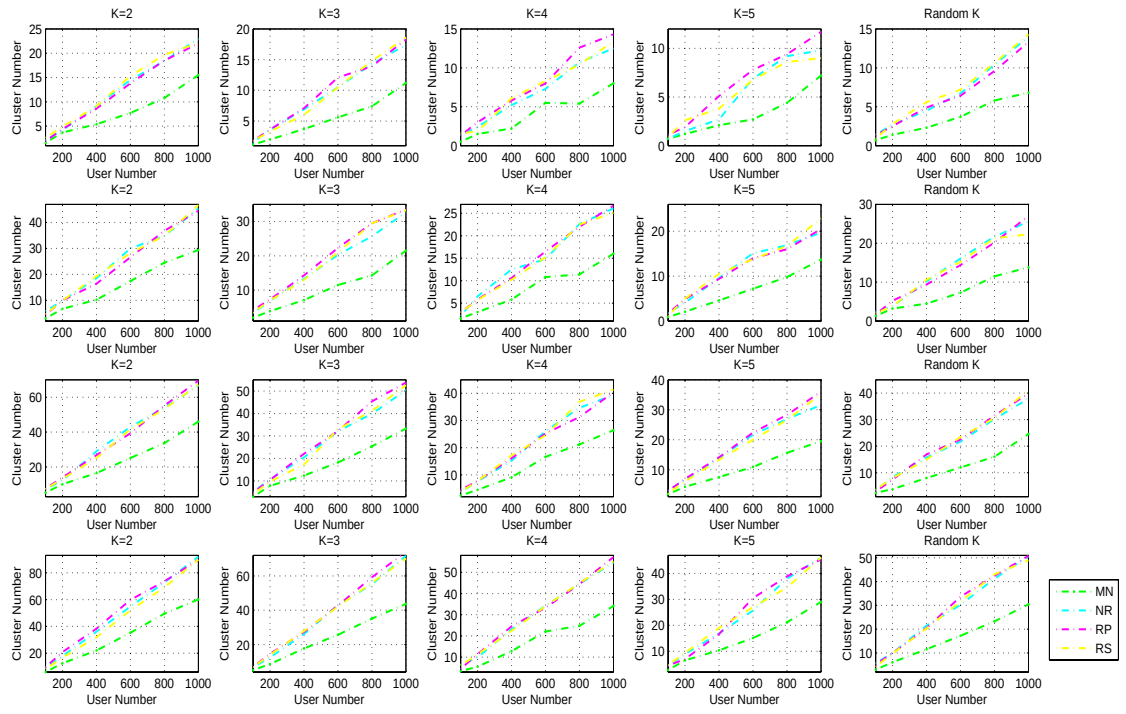

Figure 12. Cluster Robustness

This section mainly focuses on the relationship between the number of clusters adjusted and the percentage of leaving users. The less the number is, the more robust the clusters can provide. Fig.12 shows that the number of clusters adjusted is much smaller compared with the number of leaving users. Therefore, we can conclude that our ClusterCloak can provide more robustness.

### *5.4 QoS Analysis*

In this section, the relationship between $R_s = S_c / S$ and the number of users is analyzed. $S_c$ is defined as the area of a cluster C. S is defined as the total area of all clusters. The relation between $R_s$ and the cluster size is approximately linear. If TTP sends a smaller region to the LBS, a smaller list of results will be returned. Fig.13 shows $R_s$ is much lower than 1 in all the methods. Though the MBR area is small, K-anonymity can still be guaranteed. Therefore, ClusterCloak can provide accurate QoS.

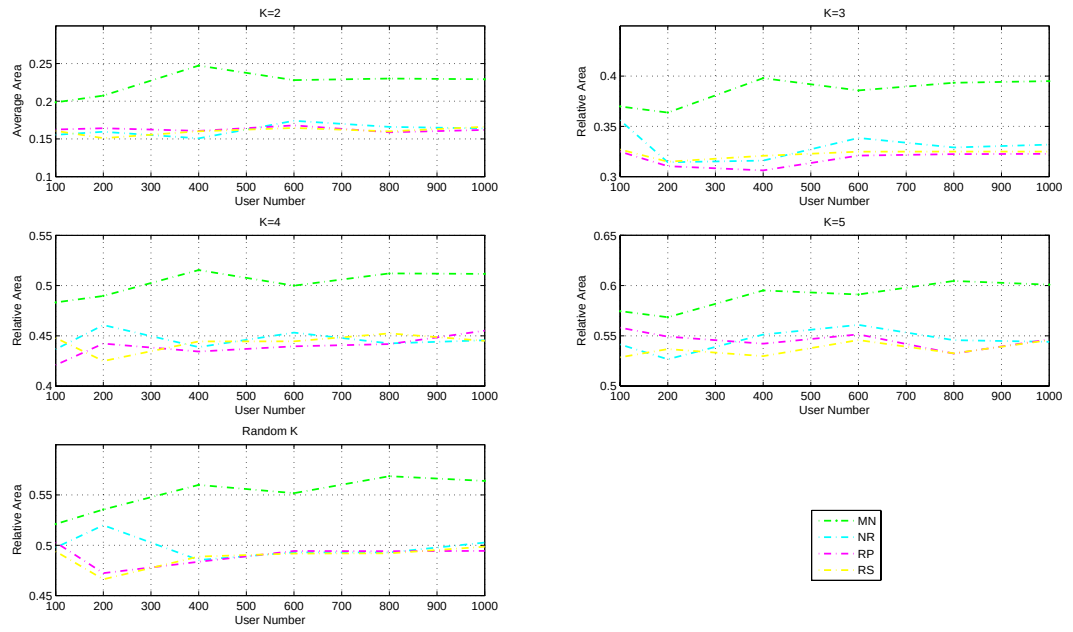

Figure 13. QoS Analysis

## VI. CONCLUSIONS

In this paper, we have proposed a location privacy preserving scheme for pervasive computing environment named ClusterCloak. Our approach can effectively protect location privacy with personalized K-anonymity while satisfying the privacy and QoS requirements of the users. ClusterCloak is adopted by TTP, and clusters can be adjusted in real time when users move from one domain to another domain. The theoretical and experimental analysis proves that our approach can provide more privacy, more accurate QoS, more robustness and lower complexity, which balances the security and the requirements of the pervasive computing devices.

Our future work will include:

- Heuristic methods: The methods of selecting initial points may affect the accuracy of ClusterCloak. Therefore, we will study the heuristic methods for ClusterCloak to achieve higher privacy level.
- MAC address privacy: ClusterCloak can only achieve location privacy for users in the application layer. Next, we will protect location privacy for MAC address.

## ACKNOWLEDGMENTS

This work was partially supported by the National Natural Science Foundation of China under Grants No.60703101 and No.60903153, and the Fundamental Research Funds for the Central Universities.